\documentclass[aps,prb,twocolumn,showpacs]{revtex4}

\usepackage{graphicx}
\usepackage{amsmath}
\usepackage{amssymb}

\begin{document}

\title{Evolution of electron Fermi surface with doping in cobaltates}

\author{Xixiao Ma}

\affiliation{Department of Physics, Beijing Normal University, Beijing 100875, China}

\author{Yu Lan}

\affiliation{College of Physics and Electronic Engineering, Hengyang Normal University, Hengyang 421002, China}

\author{Ling Qin}

\affiliation{College of Physics and Engineering, Chengdu Normal University, Chengdu 611130, China}

\author{L\"ulin Kuang}

\affiliation{Sugon National Research Center for High-performance Computing Engineering Technology, Beijng 100093, China}

\author{Shiping Feng}
\thanks{Corresponding author. E-mail: spfeng@bnu.edu.cn}

\affiliation{Department of Physics, Beijing Normal University, Beijing 100875, China~~}

\begin{abstract}
The notion of the electron Fermi surface is one of the characteristic concepts in the field of condensed matter physics, and it plays a crucial role in the understanding of the physical properties of doped Mott insulators. Based on the $t$-$J$ model, we study the nature of the electron Fermi surface in the cobaltates, and qualitatively reproduce the essential feature of the evolution of the electron Fermi surface with doping. It is shown that the underlying hexagonal electron Fermi surface obeys Luttinger's theorem. The theory also predicts a Fermi-arc phenomenon at the low-doped regime, where the region of the hexagonal electron Fermi surface along the $\Gamma$-$K$ direction is suppressed by the electron self-energy, and then six disconnected Fermi arcs located at the region of the hexagonal electron Fermi surface along the $\Gamma$-$M$ direction emerge. However, this Fermi-arc phenomenon at the low-doped regime weakens with the increase of doping.
\end{abstract}


\maketitle

\section{Introduction}\label{Introduction}

Investigation of the oxide conductors has uncovered many interesting electronic states characterized by the strong electron correlation \cite{Bednorz86,Tokura89,Takada03,Schaak03}, which include unconventional superconductivity and anomalous properties in the normal-state \cite{Bednorz86,Tokura89,Takada03,Schaak03,Kastner98,Armitage10,Wang03,Wang04,Luo04,Sales04}. Except for the cuprates \cite{Bednorz86,Tokura89}, the discovery of superconductivity in the cobaltates \cite{Takada03,Schaak03} has generated great interest, since the cobaltate is probably the only system other than the cuprates where a doped Mott insulator becomes a superconductor. However, unlike the cuprates that have a square lattice, the cobaltates have a triangular lattice \cite{Takada03,Schaak03}, where the geometric spin frustration exists. In a doped Mott insulator, a central issue in the theory concerns the nature and topology of the electron Fermi surface (EFS), since it can be measured experimentally and its shape can have crucial influence on the physical properties. The first principle band calculations predict that the cobaltates have a large EFS centered around the $\Gamma$ point of the Brillouin zone (BZ) and six small pockets near the $K$ points \cite{Singh00}. However, the angle-resolved photoemission spectroscopy (ARPES) measurements on the cobaltates reveal only the large EFS \cite{Hasan04,Yang04,Chainani04,Yang05,Shimojima05,Qian06,Qian06a,Shimojima06,Nicolaou10}, while six small EFS pockets near the $K$ points are absent. In particular, these ARPES experimental data also indicate that the area of EFS contains $1+\delta$ electrons, and therefore fulfills Luttinger's theorem \cite{Yang05,Qian06a,Nicolaou10}, where $\delta$ is the electron doping concentration. In particular, there is a remarkable resemblance in the normal- and SC-state properties between the cobaltates and cuprates \cite{Bednorz86,Tokura89,Takada03,Schaak03,Kastner98,Armitage10,Wang03,Wang04,Luo04,Sales04}, reflecting a fact that the strong electron correlation is common for both these materials \cite{Armitage10,Anderson87,Phillips10}. It has been originally argued that the strong electron correlation may destroy the one-electron band structure \cite{Anderson87}, however, the observed EFS \cite{Hasan04,Yang04,Chainani04,Yang05,Shimojima05,Qian06,Qian06a,Shimojima06,Nicolaou10} shows that a momentum-space picture also is an appropriate description of the electronic structure in a doped Mott insulator.

The understanding of the nature of EFS and its evolution with doping in a doped Mott insulator is a necessary step toward understanding of the anomalous normal-state properties of the system and the related superconducting mechanism. Although the evolution of EFS with doping in the cobaltates has been well established by now \cite{Hasan04,Yang04,Chainani04,Yang05,Shimojima05,Qian06,Qian06a,Shimojima06,Nicolaou10}, its full understanding is still a challenging issue. Within the framework of the local-spin-density approximation, the EFS topology in the cobaltates has been studied numerically by taking into account the on-site Coulomb interaction \cite{Zhang04}, however, the obtained EFS area is twice as large, and therefore is inconsistent with the ARPES observations. Based on the multiband Hubbard model, the dynamical mean-field (MF) numerical simulation indicates that the Fermi pockets in the cobaltates become even larger in size than the local density approximation \cite{Ishida05}. On the other hand, the EFS topology in a multiorbital Hubbard model on a triangular lattice has been studied within the Hartree-Fock and strong-coupling Gutzwiller approximation \cite{Zhou05,Korshunov07}, where the obtained quasiparticle dispersion and the EFS topology in the presence of the strong on-site Coulomb interaction is in qualitative agreement with the ARPES measurements on the cobaltates. In particular, the multiorbital Gutzwiller approximation \cite{Zhou05,Korshunov07} also shows that the six small pockets near the $K$ points predicted by the first principle band calculations are absent due to the strong electron correlation, which pushes the $e'_{g}$ band below the Fermi level, leading to an orbital polarized state with a single hexagonal EFS. Moreover, the specific EFS topology has been discussed within a combined cluster calculation and renormalization group approach \cite{Kiesel13}, and then some unusual properties has been explained. In our recent work \cite{Feng15b}, the nature of EFS and its evolution with doping in the cuprates has been studied by taking into account the electron self-energy effect due to the interaction between electrons by the exchange of spin excitations, and the results show clearly that although there is a large EFS satisfying Luttinger's theorem, the antinodal region of EFS is suppressed by the electron self-energy, and then the low-energy electron excitations occupy four disconnected Fermi arcs located around the nodal region, in qualitative agreement with the ARPES observations \cite{Ding97,Norman98,Yoshida09,Meng11,Comin14,Hashimoto15}. In particular, our theoretical results \cite{Feng15b} also indicate that the charge-order state in the cuprates \cite{Comin14,Hashimoto15,Comin15,Wu11,Chang12} is driven by the Fermi-arc instability, with a characteristic wave vector corresponding to the hot spots of the Fermi arcs rather than the antinodal nesting vector. In this paper, we study the nature of EFS and its evolution with doping in the cobaltates along with this line. In particular, we show that the area enclosed by the hexagonal EFS is identical to the total numbers of electrons as expected from Luttinger's theorem. Moreover, our theoretical result predicts a Fermi-arc phenomenon in the cobaltates at the low-doped regime, where the region of the hexagonal EFS along the $\Gamma$-$K$ direction is suppressed by the electron self-energy, leading to that the hexagonal EFS is broken into six disconnected Fermi arcs centered around the region along the $\Gamma$-$M$ direction.

The rest of this paper is organized as follows. We present the basic formalism in Section \ref{Formalism}, where we generalize the electron Green's function from the previous case of the doped square-lattice Mott insulators to the present case for the doped triangular-lattice Mott insulators, and then evaluate explicitly the electron spectral function. Within this theoretical framework, we therefore discuss the evolution of EFS of the cobaltates with doping in Section \ref{electronic-structure}. It is shown that the Fermi-arc phenomenon in the cobaltates at the low-doped regime weakens with the increase of doping. Finally, we give a summary in Section \ref{conclusions}.

\section{Theoretical framework}\label{Formalism}

Following the discovery of the cuprate superconductors \cite{Bednorz86,Tokura89}, there were various suggestions of way in which the strongly repulsive Hubbard model, or its equivalent, the $t$-$J$ model can capture the essential physics in a doped Mott insulator \cite{Anderson87,Phillips10,Baskaran03}. The $t$-$J$ model consists of two parts, the kinetic energy and magnetic energy parts, respectively. The kinetic-energy term in the present study includes the electron nearest-neighbor (NN) hopping term $t$ and next NN hopping term $t'$, and therefore it makes electrons itinerant. In particular, the NN hoping produces EFS, while the next NN hoping produces the curvature. On the other hand, the Heisenberg magnetic energy term with the NN spin-spin antiferromagnetic (AF) exchange $J$ describes AF coupling between localized spins. The cobaltate is an electron-doped Mott insulator \cite{Takada03,Schaak03}. In this case, the electron-doped $t$-$J$ model on a triangular lattice is very difficult to analyze, analytically as well as numerically, because of the restriction of the motion of electrons in the restricted Hilbert space without zero electron occupancy, i.e., $\sum_{\sigma}C^{\dagger}_{l\sigma} C_{l\sigma}\geq 1$, where $C^{\dagger}_{l\sigma}$ ($C_{l\sigma}$) creates (destroys) one electron with spin $\sigma$ on site $l$. In the hole-doped side, it has been shown that the local constraint of no double electron occupancy can be treated properly within the fermion-spin approach \cite{Feng15,Feng0494}. However, to apply this fermion-spin theory to the electron-doped case, we \cite{Liu04} can work in the hole representation via a particle-hole transformation $C_{l\sigma}\rightarrow f^{\dagger}_{l-\sigma}$, with $f^{\dagger}_{l\sigma}$ ($f_{l\sigma}$) that is the hole creation (annihilation) operator, and then the local constraint of no zero electron occupancy in the electron representation $\sum_{\sigma}C^{\dagger}_{l\sigma}C_{l\sigma}\geq 1$ is replaced by the local constraint of no double occupancy in the hole representation $\sum_{\sigma} f^{\dagger}_{l\sigma} f_{l\sigma}\leq 1$. This local constraint of no double occupancy now can be dealt properly by the fermion-spin theory \cite{Feng15,Feng0494}, $f_{l\uparrow}=a^{\dagger}_{l\uparrow} S^{-}_{l}$ and $f_{l\downarrow}=a^{\dagger}_{l\downarrow}S^{+}_{l}$, where the spinful fermion operator $a_{l\sigma}=e^{-i\Phi_{l\sigma}}a_{l}$ keeps track of the charge degree of freedom of the constrained hole together with some effects of spin configuration rearrangements due to the presence of the doped electron itself (charge carrier), while the spin operator $S_{l}$ represents the spin degree of freedom of the constrained hole, then the local constraint of no double occupancy is satisfied in the actual calculations. In particular, the restriction of no double occupancy in a given site induces a strong coupling between the charge and spin degrees of freedom of the constrained hole, which leads to that the spin configuration in the $t$-$J$ model is strongly rearranged due to the effect of the charge-carrier hopping on the spins. In the fermion-spin theory, the charge transport is mainly governed by the scattering of charge carriers due to spin fluctuations, while the scattering of spins due to charge-carrier fluctuations dominates the spin dynamics. Based on the triangular-lattice $t$-$J$ model in the fermion-spin representation, the charge transport of the cobaltates has been discussed \cite{Liu04}, and the obtained results show that the conductivity spectrum has a low-energy peak and an unusual midinfrared band, while the resistivity is characterized by a crossover from the high-temperature metallic-like to low-temperature insulating-like behavior, in qualitative agreement with the experimental observations \cite{Wang04}.

For the discussions of the nature of EFS and its evolution with doping in the cobaltate, we need to calculate the electron Green's function of the triangular-lattice $t$-$J$ model, which is defined as $G(l-l',t-t')=\langle\langle C_{l\sigma}(t);C^{\dagger}_{l'\sigma}(t')\rangle\rangle$. This electron Green's function in the framework of the charge-spin separation is characterized by the charge-spin recombination \cite{Anderson00,Lee99,Yu92}, i.e., the charge and spin degrees of freedom must combine to form the constrained electron. In our recent discussions of the electronic structure of the cuprates, we \cite{Feng15b} have developed a full charge-spin recombination scheme, where we have shown that in the doped square-lattice Mott insulators, the coupling form between the electron quasiparticle and spin excitation is the same as that between the charge-carrier quasiparticle and spin excitation, and then the obtained electron Green's function can give a consistent description of the electronic structure of the cuprates. Following our previous discussions for the doped square-lattice Mott insulators \cite{Feng15b}, the full electron Green's function of the $t$-$J$ model on a triangular lattice can be obtained as,
\begin{eqnarray}\label{EGF}
G({\bf k},\omega)&=&{1\over \omega-\varepsilon_{\bf k}-\Sigma_{1}({\bf k},\omega)},
\end{eqnarray}
where the MF electron excitation spectrum $\varepsilon_{\bf k}=Zt\gamma_{\bf k}-Zt'\gamma_{\bf k}'-\mu$, $\gamma_{\bf k}=[\cos k_{x}+2\cos(k_{x}/2)\cos(\sqrt{3} k_{y}/2)]/3$, $\gamma_{\bf k}'=[\cos (\sqrt{3}k_{y})+2\cos(3k_{x}/2)\cos(\sqrt{3}k_{y}/2)]/3$, and $Z$ is the number of the NN or next NN sites on a triangular lattice, while the electron self-energy $\Sigma_{1}({\bf k},\omega)$ can be evaluated in terms of the spin bubble as,
\begin{eqnarray}\label{ESE}
\Sigma_{1}({\bf k},i\omega_{n})&=&{1\over N^{2}}\sum_{{\bf p},{\bf p}'}\Lambda^{2}_{{\bf p}+{\bf p}'+{\bf k}}\nonumber\\
&\times&{1\over \beta}\sum_{ip_{m}} G({\bf p} +{\bf k},ip_{m}+i\omega_{n}) \Pi({\bf p},{\bf p}',ip_{m}),~~~~~
\end{eqnarray}
with $\Lambda_{{\bf k}}=Zt\gamma_{\bf k}-Zt'\gamma_{\bf k}'$ and the spin bubble,
\begin{eqnarray}\label{SB}
\Pi({\bf p},{\bf p}',ip_{m})&=&{1\over\beta}\sum_{ip'_{m}}D^{(0)}({\bf p'},ip_{m}')\nonumber\\
&\times&D^{(0)}({\bf p}'+{\bf p},ip_{m}'+ip_{m}),
\end{eqnarray}
where the MF spin Green's function $D^{(0)}(l-l',t-t')=\langle\langle S^{+}_{l}(t);S^{-}_{l'}(t')\rangle\rangle$ has been obtained as \cite{Liu04} $D^{(0)-1}({\bf p},\omega)=( \omega^{2}-\omega^{2}_{\bf p})/B_{\bf p}$, with the function $B_{\bf p}=\lambda_{1}[2\chi^{\rm z}_{1}(\epsilon\gamma_{\bf p}-1)+\chi_{1}(\gamma_{\bf p}-\epsilon)]-\lambda_{2} (2\chi^{\rm z}_{2}\gamma_{\bf p}'-\chi_{2})$ and the MF spin excitation spectrum,
\begin{widetext}
\begin{eqnarray}\label{MFSES}
\omega^{2}_{\bf k}&=&\lambda_{1}^{2}\left [{1\over 2}\epsilon\left (A_{1}-{1\over 2}\alpha\chi^{\rm z}_{1}-\alpha\chi_{1}\gamma_{\bf k}\right)(\epsilon-\gamma_{\bf k})+\left (A_{2} -{1\over 2Z}\alpha\epsilon\chi_{1}-\alpha \epsilon\chi^{\rm z}_{1}\gamma_{\bf k}\right )(1-\epsilon\gamma_{\bf k})\right ]+\lambda_{2}^{2}\left [\alpha\left (\chi^{\rm z}_{2} \gamma_{\bf k}'-{5\over 2Z}\chi_{2}\right )\gamma_{\bf k}'\right . \nonumber\\
&+&\left . {1\over 2}\left (A_{3}-{1\over 3}\alpha\chi^{\rm z}_{2}\right )\right ]+\lambda_{1}\lambda_{2}\left [\alpha\chi^{\rm z}_{1} (1-\epsilon\gamma_{\bf k})\gamma_{\bf k}'
+{1\over 2}\alpha(\chi_{1}\gamma_{\bf k}'-C_{3})(\epsilon-\gamma_{\bf k})+\alpha \gamma_{\bf k}'(C^{\rm z}_{3}-\epsilon\chi^{\rm z}_{2}\gamma_{\bf k})-{1\over 2}\alpha \epsilon (C_{3}-\chi_{2}\gamma_{\bf k})\right ],~~~~
\end{eqnarray}
\end{widetext}
where $\lambda_{1}=2ZJ_{\rm eff}$, $\lambda_{2}=4Z\phi_{2}t'$, $\epsilon=1+2t\phi_{1}/J_{\rm eff}$, $J_{\rm eff}= (1-\delta)^{2}J$, $\delta=\langle a^{\dagger}_{l\sigma}a_{l\sigma} \rangle=\langle a^{\dagger}_{l}a_{l}\rangle$ is the doping concentration, $A_{1}=\alpha C_{1}+(1-\alpha)/(2Z)$, $A_{2}=\alpha C^{\rm z}_{1}+(1-\alpha)/(4Z)$, $A_{3}=\alpha C_{2}+(1-\alpha) /(2Z)$, the charge-carrier's particle-hole parameters $\phi_{1}=\langle a^{\dagger}_{l\sigma}a_{l+\hat{\eta}\sigma} \rangle$ and $\phi_{2}=\langle a^{\dagger}_{l\sigma}a_{l+\hat{\tau}\sigma} \rangle$, the spin correlation functions $\chi_{1}=\langle S^{+}_{l}S^{-}_{l+\hat{\eta}}\rangle$, $\chi_{2}=\langle S^{+}_{l} S^{-}_{l+\hat{\tau}}\rangle$, $\chi^{\rm z}_{1}=\langle S_{l}^{\rm z}S_{l+\hat{\eta}}^{\rm z} \rangle$, $\chi^{\rm z}_{2} =\langle S_{l}^{\rm z}S_{l+\hat{\tau}}^{\rm z}\rangle$, $C_{1}=(1/Z^{2})\sum_{\hat{\eta},\hat{\eta'}}\langle S_{l+\hat{\eta}}^{+}S_{l+\hat{\eta'}}^{-} \rangle$, $C^{\rm z}_{1}= (1/Z^{2})\sum_{\hat{\eta},\hat{\eta'}}\langle S_{l+\hat{\eta}}^{z}S_{l+\hat{\eta'}}^{z}\rangle$, $C_{2}=(1/Z^{2})\sum_{\hat{\tau},\hat{\tau'}}\langle S_{l+\hat{\tau}}^{+} S_{l+\hat{\tau'}}^{-}\rangle$, $C_{3}=(1/Z) \sum_{\hat{\tau}}\langle S_{l+\hat{\eta}}^{+}S_{l+\hat{\tau}}^{-}\rangle$, and $C^{\rm z}_{3}=(1/Z)\sum_{\hat{\tau}} \langle S_{l+\hat{\eta}}^{\rm z}S_{l+\hat{\tau}}^{\rm z}\rangle$, with $\hat{\eta}$ and $\hat{\tau}$ that represent nearest-neighbors and next nearest-neighbors, respectively, for each site $l$. In order not to violate the sum rule of the correlation function $\langle S^{+}_{l} S^{-}_{l}\rangle=1/2$ in the case without AF long-range-order, an important decoupling parameter $\alpha$ has been introduced in the decoupling approximation for the higher order spin Green's function, which can be regarded as the vertex correction \cite{Feng15}.

In this full electron Green's function (\ref{EGF}), the dynamics of an electron quasiparticle at energy $\omega$ and momentum ${\bf k}$ is determined by the scattering processes between the electron quasiparticle itself and the internal spin degree of freedom of the electron with which the electron quasiparticle is coupled. The scattering processes renormalize the electron quasipaticle energies, and affect the lifetime. The inverse of the imaginary part of the electron self-energy $\Sigma_{1}({\bf k},\omega)$ provides the electron quasiparticle lifetime. In particular, the electron self-energy $\Sigma_{1}({\bf k},\omega)$ in Eq. (\ref{EGF}) is not an even function of $\omega$, and is self-consistently related with the electron Green's function $G({\bf k},\omega)$. For the evaluation of $\Sigma_{1}({\bf k},\omega)$, we break up $\Sigma_{1}({\bf k},\omega)$ as $\Sigma_{1}({\bf k},\omega)=\Sigma_{\rm 1e} ({\bf k},\omega)+\omega\Sigma_{\rm 1o}({\bf k},\omega)$, with $\Sigma_{\rm 1e} ({\bf k},\omega)$ and $\omega\Sigma_{\rm 1o}({\bf k},\omega)$ are the corresponding symmetric and antisymmetric parts of $\Sigma_{1}({\bf k},\omega)$, respectively, and then both $\Sigma_{\rm 1e}({\bf k}, \omega)$ and $\Sigma_{\rm 1o}({\bf k},\omega)$ are even functions of $\omega$. In the static limit, we follow the common practice \cite{Mahan81}, and define the electron quasiparticle coherent weight as $Z^{-1}_{\rm F}=1-{\rm Re} \Sigma_{\rm 1o}({\bf k},\omega=0)\mid_{{\bf k}_{\rm A}}$, where the wave vector ${\bf k}$ in $Z_{\rm F}({\bf k})$ has been chosen as ${\bf k}_{\rm A}=[{\pi/3},{{{\sqrt 3}\pi}/3}]$. In this case, the full electron Green's function in Eq. (\ref{EGF}) can be evaluated explicitly as,
\begin{eqnarray}\label{BCSEGF}
G({\bf k},i\omega_{n})&=&Z_{\rm F}\over{i\omega_{n}-\bar{\varepsilon}_{\bf k}},
\end{eqnarray}
with $\bar{\varepsilon}_{\bf k}=Z_{\rm F}\varepsilon_{\bf k}$, reflecting a fact that in the static-limit approximation, the electron quasiparticle is renormalized by a factor $Z_{\rm F}$, which therefore suppresses the spectral weight of the single-particle excitation spectrum \cite{Feng15b}. Eq. (\ref{BCSEGF}) also reflects a fact that in an interacting electron system, the elementary excitation can be described as the electron quasiparticle, but with a quasiparticle coherent weight $Z_{\rm F}$ that decreases as the electron correlation (then the interaction) increase, the remaining weight being transferred to the incoherent electron excitations. Substituting this electron Green's function (\ref{BCSEGF}) and spin bubble (\ref{SB}) into Eq. (\ref{ESE}), the electron self-energy is therefore obtained explicitly as,
\begin{eqnarray}\label{ESE1}
\Sigma_{1}({\bf k},\omega)&=&{1\over N^{2}}\sum_{{\bf pp'}\nu}(-1)^{\nu+1}Z_{\rm F}\Omega_{\bf pp'k}\left ({F^{(\nu)}_{{\rm 1} {\bf pp'k}} \over\omega+\omega_{\nu{\bf p}{\bf p}'}- {\bar{\varepsilon}}_{{\bf p}+{\bf k}}}\right . \nonumber\\
&+&\left . {F^{(\nu)}_{{\rm 2}{\bf pp'k}}\over\omega-\omega_{\nu{\bf p}{\bf p}'}-{\bar{\varepsilon}}_{{\bf p}+{\bf k}}} \right ),
\end{eqnarray}
where $\nu=1,2$, $\Omega_{\bf pp'k}=\Lambda^{2}_{{\bf p}+{\bf p}'+{\bf k}}B_{{\bf p}'}B_{{\bf p}+{\bf p}'}/(4\omega_{{\bf p}'}\omega_{{\bf p}+{\bf p}'})$, $\omega_{\nu{\bf p}{\bf p}'}=\omega_{{\bf p}+{\bf p}'}-(-1)^{\nu}\omega_{\bf p'}$, and the functions, $F^{(\nu)}_{{\rm 1}{\bf pp'k}}=n_{\rm F}({\bar{\varepsilon}}_{{\bf p}+{\bf k}})n^{(\nu)}_{{\rm 1B} {\bf pp'}}+n^{(\nu)}_{{\rm 2B}{\bf pp'}}$, $F^{(\nu)}_{{\rm 2}{\bf pp'k}}= [1-n_{\rm F}({\bar{\varepsilon}}_{{\bf p}+{\bf k}})]n^{(\nu)}_{{\rm 1B}{\bf pp'}}+n^{(\nu)}_{{\rm 2B}{\bf pp'} }$, $n^{(\nu)}_{{\rm 1B}{\bf pp'}}=1+n_{\rm B}(\omega_{{\bf p}'+{\bf p} })+n_{\rm B}[(-1)^{\nu+1}\omega_{\bf p'}]$, $n^{(\nu)}_{{\rm 2B}{\bf pp'}}=n_{\rm B}(\omega_{{\bf p}'+{\bf p}} ) n_{\rm B}[(-1)^{\nu+1} \omega_{\bf p'}]$, while $n_{\rm B}(\omega)$ and $n_{\rm F}(\omega)$ are the boson and fermion distribution functions, respectively. In this case, the electron quasiparticle coherent weight $Z_{\rm F}$ satisfies the self-consistent equation,
\begin{eqnarray}\label{ESCE1}
{1\over Z_{\rm F}}&=&1+{1\over N^{2}}\sum_{{\bf pp'}\nu}(-1)^{\nu+1}Z_{\rm F}\Omega_{{\bf pp'}{\bf k}_{\rm A}}
\left ({F^{(\nu)}_{{\rm 1}{\bf pp}'{\bf k}_{\rm A}}\over (\omega_{\nu{\bf p}{\bf p}'}-{\bar{\varepsilon}}_{{\bf p}+{\bf k}_{\rm A}})^{2}}\right .\nonumber\\
&+&\left . {F^{(\nu)}_{{\rm 2}{\bf pp}'{\bf k}_{\rm A}}\over (\omega_{\nu{\bf p}{\bf p}'}+ {\bar{\varepsilon}}_{{\bf p}+{\bf k}_{\rm A}})^{2}}\right ). ~~~~~
\end{eqnarray}
This equation (\ref{ESCE1}) must be solved simultaneously with the self-consistent equation,
\begin{eqnarray}\label{ESCE2}
1+\delta &=& {1\over N}\sum_{{\bf k}}Z_{\rm F}\left (1-{\rm tanh}[{1\over 2}\beta\bar{\varepsilon}_{\bf k}] \right ),
\end{eqnarray}
and then the electron quasiparticle coherent weight $Z_{\rm F}$ and chemical potential $\mu$ are determined self-consistently.

\section{Quantitative characteristics}\label{electronic-structure}

We are now ready to discuss the nature of EFS and its evolution with doping in the cobaltates. In the cobaltates, it has been shown experimentally that the value of the effective electron hopping $t$ is on the order of the magnetic exchange coupling $J$ \cite{Hasan04,Yang04}, and in this case, the commonly used parameters in this paper are chosen as $t/J=1.5$ and $t'/t=0.2$ for a qualitative discussion. Unless specified, the lattice parameter and reduced Planck constant are set to unity.

\subsection{Evolution of electron Fermi surface with doping}\label{electron-spectrum}

\begin{figure*}[t!]
\centering
\includegraphics[scale=0.5]{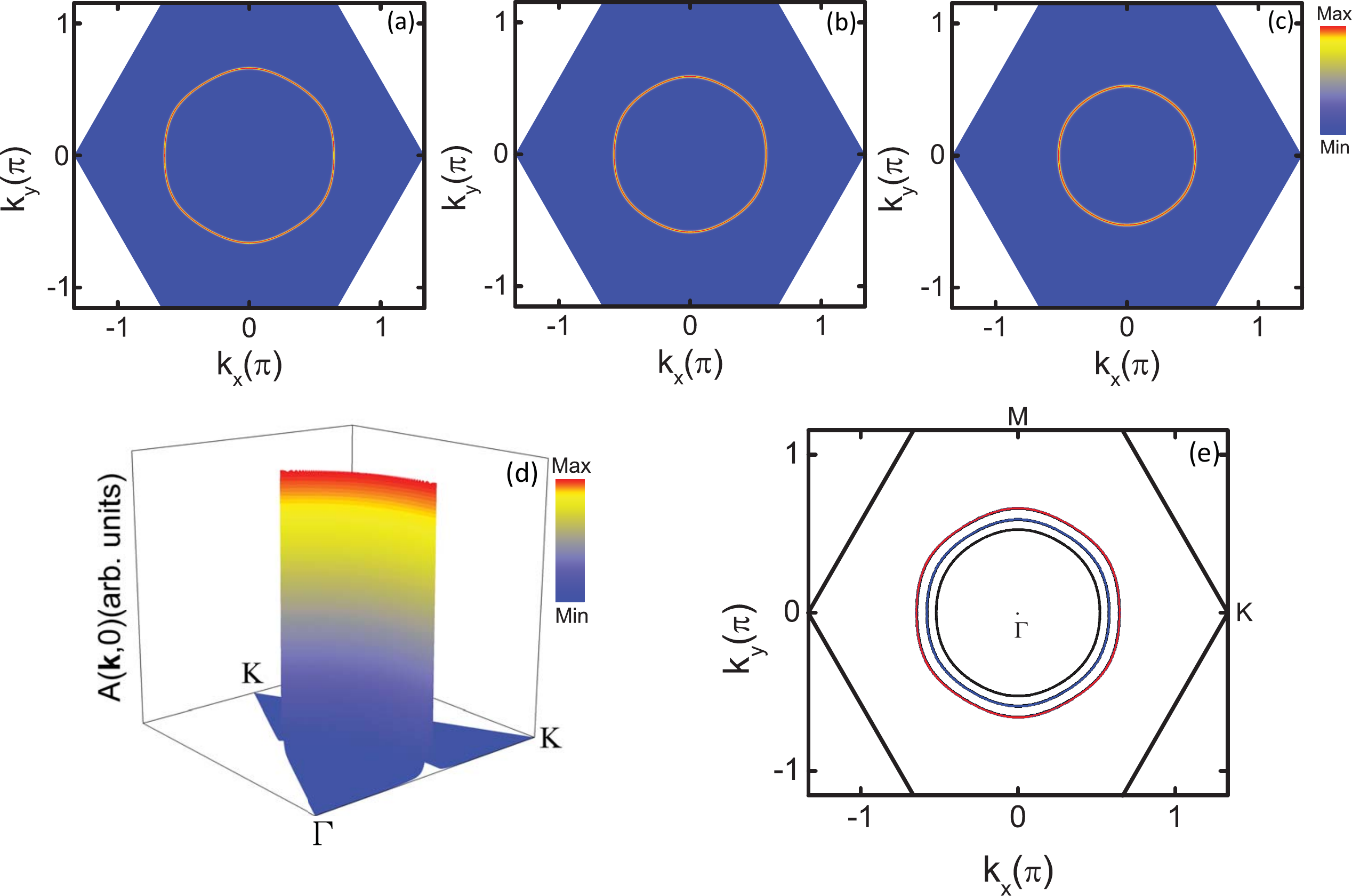}
\caption{(Color online) The map of the spectral intensity $A({\bf k},0)$ at (a) $\delta=0.32$, (b) $\delta=0.40$, and (c) $\delta=0.45$ with $T=0.001J$ for $t/J=1.5$ and $t'/t=0.2$ in the Brillouin zone. (d) The electron spectral function $A({\bf k},0)$ in the $[k_{x},k_{y}]$ plane at $\delta=0.32$. (e) EFS locations at $\delta=0.32$ (red line), $\delta=0.40$ (blue line), and $\delta=0.45$ (black line) in the Brillouin zone.
\label{spectral-maps}}
\end{figure*}

In the framework of the fermion-spin theory, the quasiparticles of a doped Mott insulator consist of bare electrons dressed by spin excitations, and are parameterized by the electron spectral function $A({\bf k},\omega)$. In particular, the intensity of the electron spectra at zero energy is used to map out the underlying EFS. Since the low-energy electron excitations is determined by the electronic structure near EFS, the EFS topology plays a significant role in comprehending the unconventional physics in a doped Mott insulator. For a clear illustration of the effect of the energy and momentum dependence of the electron self-energy on EFS in the cobaltates, we first discuss EFS in the static-limit approximation for the electron self-energy $\Sigma_{1}({\bf k},\omega)$. In this case, the electron spectral function $A({\bf k},\omega)=-2{\rm Im}G({\bf k},\omega)$ is obtained directly from the above electron Green's function (\ref{BCSEGF}) as,
\begin{eqnarray}\label{spectrum}
A({\bf k},\omega)&=&2\pi Z_{\rm F}\delta(\omega-\bar{\varepsilon}_{\bf k}),
\end{eqnarray}
where the energy and momentum dependence in the electron self-energy $\Sigma_{1}({\bf k},\omega)$ has been dropped, and then the electron spectral function in Eq. (\ref{spectrum}) has a Dirac delta function form on the quasiparticle dispersion curves, $\bar{\varepsilon}_{\bf k}$ versus ${\bf k}$, reflecting that the quasiparticles are similar to {\it free} electrons, but weighted by the electron quasiparticle coherent weight $Z_{\rm F}$. In other words, the form of the electron spectral function in Eq. (\ref{spectrum}) is similar to that obtained in the MF approximation, and then EFS is necessarily a surface in momentum-space on which the electron lifetime becomes infinitely long in the limit as one approaches EFS. In Fig. \ref{spectral-maps}, we plot a map of the spectral intensity $A({\bf k},0)$ in Eq. (\ref{spectrum}) at the doping levels (a) $\delta=0.32$, (b) $\delta=0.40$, and (c) $\delta=0.45$ with temperature $T=0.001J$ in the hexagonal BZ, where the calculated EFS forms a continuous contour in momentum space, and then the EFS location agrees with a large EFS observed from the ARPES experiments \cite{Hasan04,Yang04,Chainani04,Yang05,Shimojima05,Qian06,Qian06a,Shimojima06,Nicolaou10}. In particular, EFS has a hexagonal shape at the low doping, and exhibits nearly straight sections perpendicular to the high symmetry directions. However, this hexagonal character at the low-doped regime weakens with the increase of doping. To see the low-energy excitations more clearly, we plot $A({\bf k},0)$ in the $[k_{x},k_{y}]$ plane at $\delta=0.32$ in Fig. \ref{spectral-maps}d, where we see that the peaks with the same height distribute uniformly along the hexagonal EFS, and then a large EFS is formed as a closed contour of the gapless excitations in momentum space, i.e., the electron spectrum is gapped everywhere except on EFS. Moreover, according to one of the self-consistent equations (\ref{ESCE2}), EFS with the area contains $1+\delta$ electrons, and therefore is consistent with that predicted by the Luttinger's theorem. Furthermore, as shown in Fig. \ref{spectral-maps}e, the area of the hexagonal EFS increases with the increase of the doping concentration, in good agreement with the ARPES experimental results \cite{Hasan04,Yang04,Chainani04,Yang05,Shimojima05,Qian06,Qian06a,Shimojima06,Nicolaou10}.

\subsection{Fermi arcs induced by electron self-energy}\label{self-energy-electronic-structure}

\begin{figure*}[t!]
\centering
\includegraphics[scale=0.2]{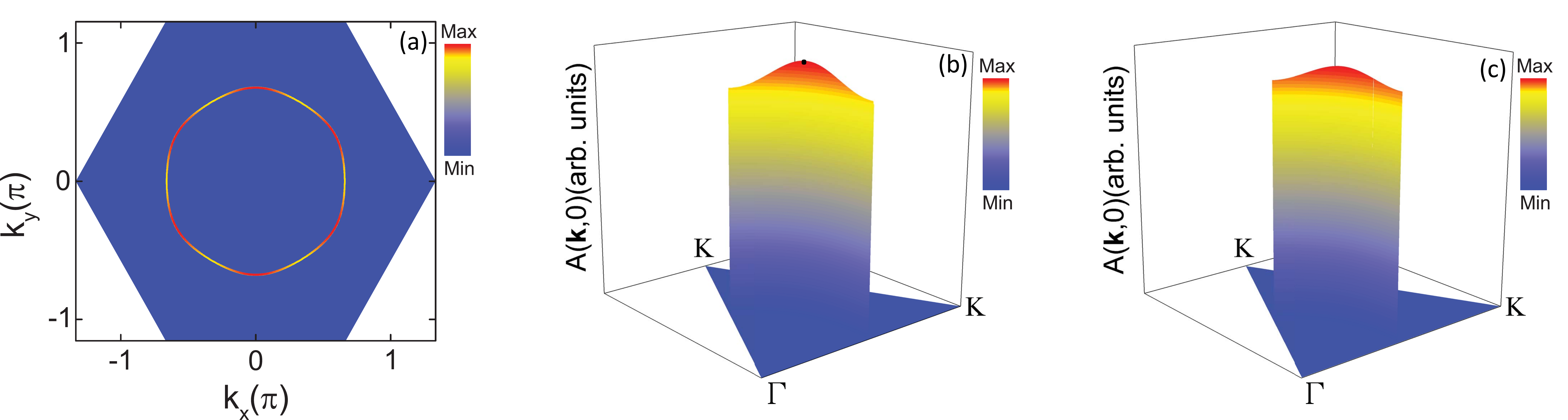}
\caption{(Color online) (a) The map of the spectral intensity $A({\bf k},0)$ at $\delta=0.32$ with $T=0.001J$ for $t/J=1.5$ and $t'/t=0.2$. (b) The electron spectral function $A({\bf k},0)$ in the $[k_{x},k_{y}]$ plane at $\delta=0.32$. The black circle indicates the location of the hot spot on the Fermi arc. (c) The electron spectral function $A({\bf k},0)$ in the $[k_{x},k_{y}]$ plane at $\delta=0.40$. \label{SE-spectral-maps}}
\end{figure*}

Now we turn to discuss the effect of the energy and momentum dependence of $\Sigma_{1}({\bf k},\omega)$ on EFS in the cobaltates. In this case, the electron spectral function can be evaluated directly from the electron Green's function (\ref{EGF}) as,
\begin{eqnarray}\label{full-spectral-function}
A({\bf k},\omega)={2|{\rm Im}\Sigma_{1}({\bf k},\omega)|\over [\omega-\varepsilon_{\bf k}-{\rm Re}\Sigma_{1}({\bf k},\omega)]^{2}+[{\rm Im}\Sigma_{1}({\bf k},\omega)]^{2}},~~~
\end{eqnarray}
where ${\rm Im}\Sigma_{1}({\bf k},\omega)$ and ${\rm Re}\Sigma_{1}({\bf k},\omega)$ are the corresponding imaginary and real parts of $\Sigma_{1}({\bf k},\omega)$ in Eq. (\ref{ESE1}), respectively. Since the strong coupling of electrons and spin excitations is considered in the electron spectral function (\ref{full-spectral-function}), the electron quasiparticle energies are renormalized, therefore the electron quasiparticles acquire a finite lifetime and the incoherent side band develops. In Fig. \ref{SE-spectral-maps}a, we plot a map of the spectral intensity $A({\bf k},0)$ in Eq. (\ref{full-spectral-function}) at $\delta=0.32$ with $T=0.001J$. Comparing it with Fig. \ref{spectral-maps}a for the same set of parameters except for the effect from the momentum and energy dependence of $\Sigma_{1}({\bf k},\omega)$, we see that the Fermi arcs emerge, where EFS around the region along the $\Gamma$-$K$ direction has been suppressed by $\Sigma_{1}({\bf k},\omega)$, leaving behind six disconnected Fermi arcs centered around the region of EFS along the $\Gamma$-$M$ direction. To see this Fermi-arc phenomenon more clearly, we plot $A({\bf k},0)$ in the $[k_{x},k_{y}]$ plane at $\delta=0.32$ in Fig. \ref{SE-spectral-maps}b, where the location of the hot spot, marked by the black circle, is thus determined by the highest peak height on the Fermi arc. Our result in Fig. \ref{SE-spectral-maps}b shows that the Fermi arc spreads around the hot spot, while these hot spots on the Fermi arcs, reflecting a strong charge fluctuation, connected by the wave vector contribute effectively to the electron quasiparticle scattering process. However, this Fermi-arc phenomenon weakens with the increase of doping, which leads to a crossover from the strong charge fluctuation at the low-doped regime to the weak charge fluctuation at the high-doped regime. In order to show this crossover clearly, we plot $A({\bf k},0)$ in the $[k_{x},k_{y}]$ plane at $\delta=0.40$ in Fig. \ref{SE-spectral-maps}c. Comparing it with Fig. \ref{SE-spectral-maps}b for the same set of parameters except for doping, we find that EFS around the region along the $\Gamma$-$K$ direction is reduced very slightly, and then the weight of the hot spot on the Fermi arc at $\delta=0.40$ is much smaller than that at $\delta=0.32$.

The cobaltates have a triangular lattice, where the spin configuration arrangement makes it impossible to minimize the energy on all spins on the lattice sites at the same time - a situation known as the geometrical spin frustration. However, this triangular-lattice structure induces possibly a geometrical construct for $120^{\circ}$-type collective instability. In particular, the earlier calculation \cite{Motrunich04} based on the extended Hubbard model on a triangular lattice has predicted that the NN repulsion can drive the commensurate $\sqrt{3}\times\sqrt{3}$ charge-order state near the commensurate doping $\delta=1/3$. This $\sqrt{3}\times\sqrt{3}$ charge order arises from weak spatial modulations of the charge density, and therefore is closely related to the low-energy electronic structure. In particular, the ARPES experiments \cite{Yang05,Qian06} have provided the evidence to support this prediction. In our present study, although the NN repulsion is not included in the $t$-$J$ model, BZ of the $\sqrt{3}\times\sqrt{3}$ charge order near the commensurate doping $\delta=1/3$ coincides with the topology of our calculated EFS. In this case, it is also possible that the $\sqrt{3}\times\sqrt{3}$ charge-order state is driven by the Fermi-arc instability. These and the related issues are under investigation now.

\subsection{Electron momentum distribution}\label{Electron-momentum-distribution}

\begin{figure}[h!]
\centering
\includegraphics[scale=0.18]{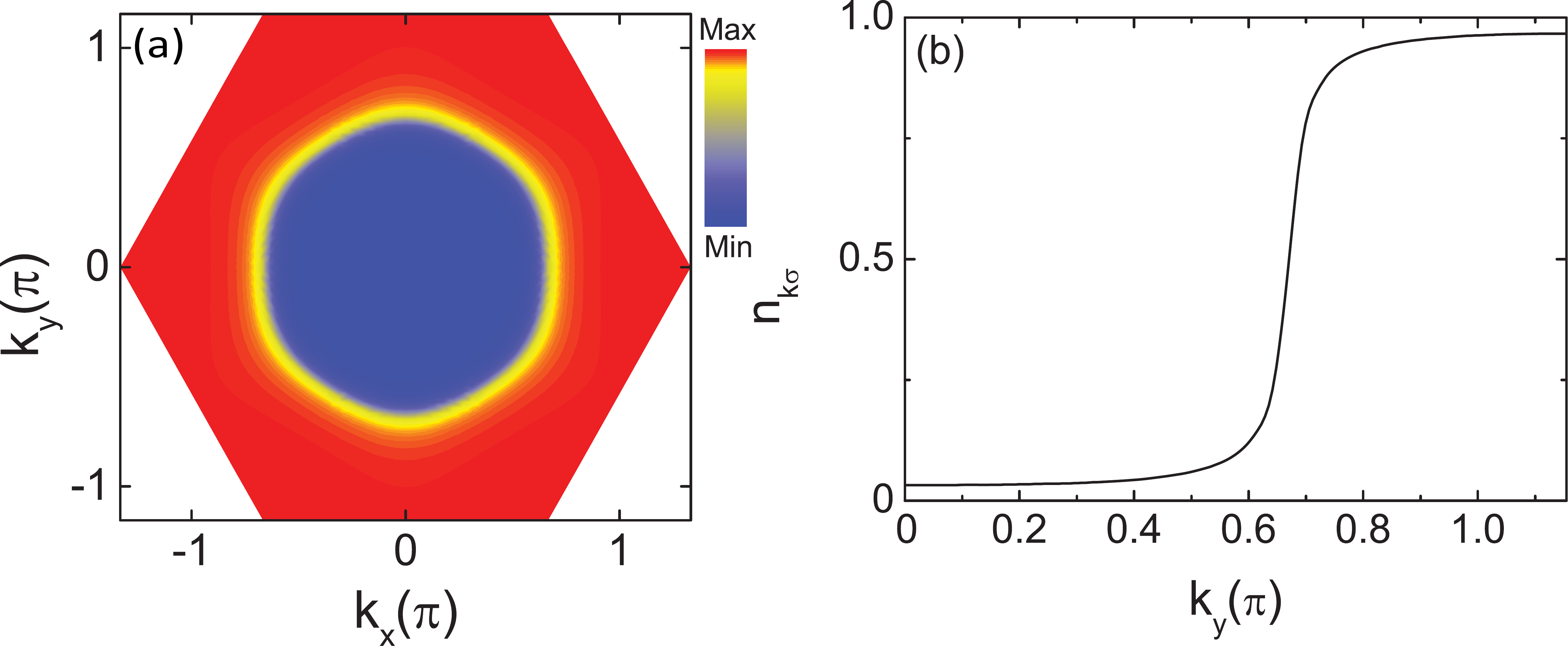}
\caption{(Color online) (a) The map of the electron-momentum distribution in the Brillouin zone at $\delta=0.32$ with $T=0.001J$ for $t/J=1.5$ and $t'/t=0.2$. (b) The electron-momentum distribution along the $\Gamma$-$M$ direction. \label{map-SE-momentum-distribution}}
\end{figure}

A quantity which is closely related to the electron spectral function (\ref{full-spectral-function}) is the electron momentum distribution,
\begin{eqnarray}\label{number}
n_{{\bf k}\sigma}= \int^{\infty}_{-\infty}{{\rm d}\omega\over 2\pi}n_{\rm F}(\omega)A({\bf k},\omega).
\end{eqnarray}
In Fig. \ref{map-SE-momentum-distribution}a, we plot a map of $n_{{\bf k}\sigma}$ indicating a shape of the hexagonal EFS in the BZ at $\delta=0.32$ with $T=0.001J$, where the identification of a large hexagonal EFS is unambiguous. To further analyze the nature of EFS, we plot $n_{{\bf k}\sigma}$ along the $\Gamma$-$M$ direction in Fig. \ref{map-SE-momentum-distribution}b. It is shown clearly that the shape of the electron momentum distribution in the cobaltates is a should-be electron momentum distribution, i.e., in some part (below the electron Fermi energy) the distribution in the presence of the interaction is closer to $1$, while in other part (above the electron Fermi energy) it is approximately closer to zero. In particular, although the line shape in Fig. \ref{map-SE-momentum-distribution}b itself is strongly dependent on the interaction, the integrated area under the curve is equal to $1+\delta$, and therefore the underlying EFS still fulfills Luttinger's theorem, which follows a fact that this integrated area is independence of the interaction.

\begin{figure}[h!]
\centering
\includegraphics[scale=0.18]{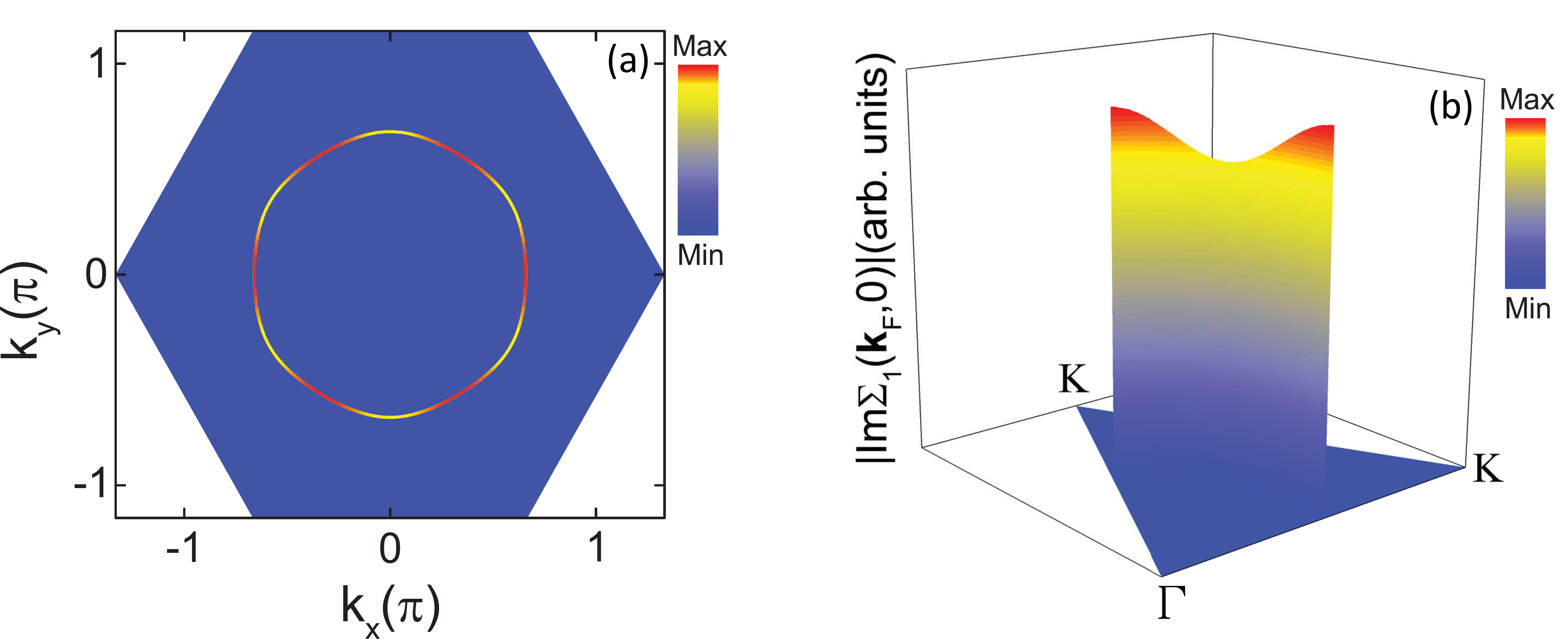}
\caption{(Color) (a) The map of $|{\rm Im}\Sigma_{1}({\bf k}_{\rm F},0)|$ at $\delta=0.32$ with $T=0.001J$ for $t/J=1.5$ and $t'/t=0.2$. (b) $|{\rm Im}\Sigma_{1}({\bf k}_{\rm F},0)|$ in the $[k_{x},k_{y}]$ plane. \label{imaginary-part-self-energy}}
\end{figure}

The essential physics of the electron quasiparticle excitations in the cobaltates is the same as in the case of the cuprates \cite{Feng15b} except for the geometric construct, and can be also attributed to the energy and momentum dependence of the electron self-energy $\Sigma_{1}({\bf k},\omega)$ in Eq. (\ref{ESE}). EFS is determined directly by $\varepsilon_{\bf k}+{\rm Re}\Sigma_{1}({\bf k},0)=0$ in the electron spectral function $A({\bf k},\omega)$ in Eq. (\ref{full-spectral-function}) at zero energy, and then the lifetime of the electron quasiparticle at EFS is determined by the inverse of the imaginary part of the electron self-energy $1/|{\rm Im}\Sigma_{1}({\bf k}_{\rm F},0)|$. In Fig. \ref{imaginary-part-self-energy}a, we map the intensity of $|{\rm Im}\Sigma_{1}({\bf k}_{\rm F},0)|$ at $\delta=0.32$ with $T=0.001J$, where ${\rm Im}\Sigma_{1}({\bf k}_{\rm F},0)$ is strongly momentum dependent. To see this point clearly, we plot $|{\rm Im}\Sigma_{1}({\bf k}_{\rm F},0)|$ in the $[k_{x},k_{y}]$ plane in Fig. \ref{imaginary-part-self-energy}b, where $|{\rm Im}\Sigma_{1}({\bf k}_{\rm F},0)|$ has a strong angular dependence with actual maximums around the region at EFS along the $\Gamma$-$K$ direction, which leads to that the part of EFS around the region along the $\Gamma$-$K$ direction is suppressed, leaving behind six disconnected segments around the region along the $\Gamma$-$M$ direction.

In the framework of the fermion-spin theory, the interaction between electrons by the exchange of spin excitations induces the electron self-energy $\Sigma_{1}({\bf k},\omega)$. As in the cuprates \cite{Feng15b,Feng12}, $\Sigma_{1}({\bf k},\omega)$ in the cobaltates is also directly related to the pseudogap $\bar{\Delta}_{\rm PG}({\bf k})$ as
$\Sigma_{1}({\bf k},\omega)\approx [\bar{\Delta}_{\rm PG}({\bf k})]^{2}/(\omega+\varepsilon_{0{\bf k}})$. This pseudogap is therefore identified as being a region of the electron self-energy effect in which the pseudogap suppresses the low-energy spectral weight of the quasiparticle excitation spectrum. In other words, the Fermi-arc phenomenon at the low-doped regime is manifested within the pseudogap regime, then the Fermi arcs and pseudogap in the cobaltates are intimately related each other, and they have a root in common originated from the electron self-energy $\Sigma_{1}({\bf k},\omega)$. However, due to the presence of the geometric spin frustration, the strength of the interaction between electrons by the exchange of spin excitations in the cobaltates is much weaker than that in the cuprates, which leads to the pseudogap parameter $\bar{\Delta}_{\rm PG}$ in the cobaltates is much smaller than that in the cuprates. As a consequence, the Fermi-arc phenomenon in the cobaltates even at the low-doped regime is not particularly obvious than that in the cuprates. On the other hand, $\bar{\Delta}_{\rm PG}$ has the same doping dependence as that discussed previously in the case of the cuprates \cite{Feng15b,Feng12}, i.e., the pseudogap state is relatively obvious at the low-doped regime, and then $\bar{\Delta}_{\rm PG}$ smoothly decreases upon increasing doping. This doping dependence of $\bar{\Delta}_{\rm PG}$ therefore leads to that the Fermi-arc phenomenon in the cobaltates at the low-doped regime weakens with the increase of doping.

Finally, we have noted that the strong-coupling single-band model lacks for a description of the $e'_{g}$ small pockets near the $K$ points. The conduction band of the cobaltates mainly consists of Co $t_{2g}$ orbitals which are split into $a_{1g}$ singlet and $e'_{g}$ doublet due to the presence of a trigonal crystal field \cite{Shimojima06}. As we have mentioned in Introduction section \ref{Introduction}, the numerical results from the first principle band calculations \cite{Singh00,Lee04} indicate the cobaltates to have a large EFS centered at the $\Gamma$ point with mainly $a_{1g}$ character and six small pockets near the $K$ points with the mostly $e'_{g}$ character for a wide range of doping. However, the first ARPES measurement on the cobaltates observed a single hexagonal $a_{1g}$ EFS only without the evidence of the small pockets \cite{Hasan04}. The preceding ARPES experimental results confirmed that the top postion of the $e'_{g}$ band is doping independent, sinking below the Fermi energy, suggesting that {\it it plays no role in the low-energy phenomena} \cite{Yang04,Chainani04,Yang05,Shimojima05,Qian06,Qian06a,Shimojima06,Nicolaou10}. The combined these ARPES experimental results \cite{Yang04,Chainani04,Yang05,Shimojima05,Qian06,Qian06a,Shimojima06,Nicolaou10} and the transport experimental data \cite{Armitage10,Wang03,Wang04,Luo04,Sales04} therefore show that {\it the essential low-energy physics} of the cobaltates is dominated by the strong electron correlation. The strong-coupling Hubbard model and its equivalent, the $t$-$J$ model, are prototypes to study the strong correlation effects in solids, especially in connection with the unconventional superconductivity \cite{Anderson87,Phillips10}. It is widely believed that the essential low-energy physics of the cuprates is contained in the single-band $t$-$J$ model on a square lattice \cite{Anderson87,Phillips10}. Although the single-band $t$-$J$ model on a triangular lattice can not be regarded as a complete model for the quantitative comparison with the cobaltates, the qualitative agreement between the present theoretical results and ARPES experimental data also indicates that the single-band $t$-$J$ model on a triangular lattice captures the essential low-energy physics of the cobaltates \cite{Baskaran03}.

\section{Conclusions}\label{conclusions}

Within the $t$-$J$ model on a triangular lattice, we have discussed the nature of EFS in the cobaltates, and qualitatively reproduce the essential feature of the evolution of EFS with doping. Our results show that the underlying hexagonal EFS satisfies Luttinger's theorem. Our theory also predicts a Fermi-arc phenomenon at the low-doped regime, where the region of the hexagonal EFS along the $\Gamma$-$K$ direction is gapped by the pseudogap, and then six disconnected Fermi arcs located at the region of the hexagonal EFS along the $\Gamma$-$M$ direction appear, which should be verified by further experiments. In particular, this Fermi-arc phenomenon at the low-doped regime weakens with the increase of doping. Incorporating the present result with that obtained in the cuprates, it is thus shown that as a natural consequence of the strong electron correlation, the Fermi-arc phenomenon may be a universal phenomenon in a doped Mott insulator in spite of the existence of a geometrical construct or not.

\section*{Acknowledgments}

The authors would like to thank Huaisong Zhao and Deheng Gao for helpful discussions. XM, LQ, LK, and SF are supported by the funds from the Ministry of Science and Technology of China under Grant No. 2012CB821403, and the National Natural Science Foundation of China (NSFC) under Grant Nos. 11274044 and 11574032. YL was supported by the Science Foundation of Hengyang Normal University under Grant No. 13B44, and Hunan Provincial Natural Science Foundation of China under Grant No. 2015JJ3027.

\end{document}